\def\QCD{}
\def\BR{\text{BR}}
\global\long\def\order#1{\mathcal{O}\left(#1\right)}
\def\Klt{K_{l3}}

\def\Vud{V_{\text{ud}}}
\def\Vus{V_{\text{us}}}
\def\Vub{V_{\text{ub}}}

\def\mKL{m_{K_L}}
\def\mpi{m_{\pi^+}}
\documentclass[letterpaper,english,prd,preprintnumbers,nofootinbib,showpacs,shortbibliography,notitlepage]{revtex4-1} %linenumbers
\usepackage[utf8]{inputenc}
\usepackage{amsmath}
\usepackage{nicefrac}

\begin{document}
\preprint{Alberta Thy 13-19}
%\title{Pion Beta Decay and CKM Unitarity}
\title{Pion beta decay and Cabibbo-Kobayashi-Maskawa unitarity}

\author{Andrzej Czarnecki }
\affiliation{Department of Physics, University of Alberta, Edmonton,  Alberta, Canada T6G 2E1}
\author{William J.~Marciano}
\affiliation{Department of Physics, Brookhaven National Laboratory, Upton, New York 11973, USA}
\author{Alberto Sirlin}
\affiliation{Department of Physics, New York University,\\ 726 Broadway, New York, New York 10003, USA}

\begin{abstract}
Pion beta decay, $\pi^+\to \pi^0 e^+\nu(\gamma)$, provides a theoretically clean $\pm 0.3\%$ determination of the CKM matrix element $\Vud$. Although impressive, that result falls short of super-allowed nuclear beta decays where an order of magnitude better precision already exists. Here, we advocate a new strategy for utilizing pion beta decay, based on its utility in determining ${\Vus / \Vud}$ via the ratio 
$R_V=\Gamma\left(K\to \pi l \nu(\gamma)\right)/\Gamma\left(\pi^+\to \pi^0 e^+\nu(\gamma)\right)$ which provides a measure of 
$  { f_+^K(0) |\Vus| / f_+^\pi(0) |\Vud| }$  independent of the Fermi constant and short-distance radiative corrections. Its dependence on the ratio of two hadronic vector current form factors provides an interesting computational goal for lattice gauge theory studies. 
Employing a recent lattice based value ${ f_+^K(0) / f_+^\pi(0) }  = 0.970(2)$, we find
$\Vus/\Vud=0.22910(91)$ compared to $\Vus/\Vud=0.23131(45)$
obtained from $R_A=\Gamma(K\to \mu\nu(\gamma))/\Gamma(\pi\to \mu\nu(\gamma))$.  
Those independent $\Vus/\Vud$ determinations exhibit a 2.2$\sigma$
discrepancy. That tension suggests: a needed shift in the lattice $ { f_+^K(0) / f_+^\pi(0) } $ value towards the consistency range 0.961(4), experimental/theory input changes or ``new physics'' effects. Other features and implications of $R_V$ and $R_A$ are also discussed.
\end{abstract}
\maketitle

\section{Introduction}
 The Standard Model (SM) of elementary particle physics includes a three-generation
 Cabibbo-Kobayashi-Maskawa (CKM) \cite{Cabibbo:1963yz,Kobayashi:1973fv} $3\times 3$ quark mixing matrix, $V_{ij}$, $i=u,c,t$, $j=d,s,b$, which
 satisfies unitarity, $V^\dagger = V^{-1}$.  That condition gives rise to
 orthonormal relationships among its rows and columns.  Of special
 interest is the first row constraint
\begin{equation}
     |\Vud|^2+|\Vus|^2+|\Vub|^2=1   .  \label{eq1}              
\end{equation}      
        Neglecting the tiny $|\Vub|^2 \simeq 2\times 10^{-5}$ contribution \cite{Tanabashi:2018oca}, it simplifies to approximately the
      original Cabibbo \cite{Cabibbo:1963yz} two-generation relationship
      \begin{equation}
             |\Vud|^2+|\Vus|^2=1    . \label{eq2}  
      \end{equation}
      An unambiguous experimental deviation from eq.~\eqref{eq2} would
      signal the presence of ``new physics'' beyond SM (BSM) expectations. BSM examples include right-handed weak currents, charged Higgs scalars, leptoquarks, $Z^\prime$ box diagram loop effects, exotic muon decay amplitudes that effectively modify the value of the Fermi constant etc. Such a discovery would have major repercussions; but, first, its acceptance would require evidence of high significance 
      (5 or more sigma effects) as well as confirmation by other measurements.
      
      Up until recently, eq.~\eqref{eq2} appeared to be well satisfied \cite{Tanabashi:2018oca,Hardy:2016vhg}.
      However, a novel dispersion relation (DR) approach \cite{Seng:2018yzq} to super-allowed nuclear beta decay loop effects found an
      increase in the electroweak radiative corrections that reduced
      $\Vud$ from $0.97420(21)$ to
\begin{equation}
     |\Vud|=0.97370(10)_\text{NP}(10)_\text{RC}   \quad \text{DR Result \cite{Seng:2018yzq}} , \label{eq3}  
\end{equation} where NP (nuclear physics) and RC  (radiative corrections) label the uncertainty sources.
     A subsequent calculation \cite{Czarnecki:2019mwq} based on a somewhat different
     approach found \begin{equation}
            |\Vud|=0.97389(10)_\text{NP}(14)_\text{RC}   \quad \text{CMS \cite{Czarnecki:2019mwq}}.  \label{eq4} 
\end{equation}
     When used with the $\Vus$ average \cite{Tanabashi:2018oca} from $K_{l3}$ ($K\to\pi l\nu$) and $K_{l2}$  ($K^\pm \to l^\pm \nu$)  decays
\begin{equation}
            |\Vus|=0.2245(8)  \quad \text{Error scale factor 2} \label{eq5},
\end{equation}
the two approaches lead to roughly 3 and 2 $\sigma$ deviations from unitarity
respectively.  Those discrepancies could be early hints of ``new physics" starting to show up as an effective deviation from CKM unitarity. Alternatively, missing nuclear physics effects and
theoretical uncertainties might eventually resolve the problem \cite{Seng:2018qru} more conventionally.

 The current $\Vus$ situation requires some explanation. $K_{l3}$  decays considered alone give a relatively small
     \begin{equation}
             |\Vus|=0.2234(8)    \quad \text{$K_{l3}$ decays}.
     \end{equation} In contrast,
 the ratio \cite{Marciano:2004uf}
     \begin{equation}
       R_A = {\Gamma(K\to \mu\nu(\gamma))\over \Gamma(\pi\to
         \mu\nu(\gamma))}, \end{equation} is generally considered a more dependable constraint, since common uncertainties in kaon and pion decays tend to cancel in the ratio. That is  particularly important for lattice gauge theory input calculations of ${f_{K^+} / f_{\pi^+}}$.
     From the experimental constraint $R_A=1.3367(28)$, one finds
        \begin{equation}
               {|\Vus|f_{K^+}\over |\Vud|f_{\pi^+}} =0.2760(4).
               \end{equation}
         Then using the lattice \QCD value \cite{Aoki:2019cca}
               \begin{equation}
                   {f_{K^+}\over f_{\pi^+}}=1.1932(19),
               \end{equation} 
    one obtains
          \begin{equation}
          \label{eq:X}
               {   |\Vus|\over |\Vud|} =0.23131(45).
               \end{equation} 
    Agreement with three-generation unitarity requires   
         \begin{equation}
           \label{eq:11}
         |\Vud|=0.97428(10)  \text{ and }   |\Vus|=0.2253(4).  %\quad  |\Vud|^2 + |\Vus|^2 =  0.99998(34) \simeq 1.
         \end{equation}
     Those SM expectations are 2 or more $\sigma$ different from some of the current $|\Vud|$  and  $|\Vus|$ values shown above.   
     $R_A$, being rather free of
     theoretical and experimental uncertainties, currently represents our best first row CKM constraint and should be taken
     seriously.
     
     Having set the stage for CKM inconsistencies we now introduce and examine a weak vector current analog of $R_A$ for which short-distance electroweak and some part of the QED radiative corrections as well as muon lifetime normalization dependence via the Fermi constant cancel in the ratio. Our new vector current analog of $R_A$ may also provide a useful target for testing lattice gauge theory calculations.
     The specific ratio we first consider is
     \begin{equation}
       \label{eq:RV}
     R_V= {  \Gamma(K_L \to\pi^\pm e^\mp\nu(\gamma))\over\Gamma(\pi^+ \to
       \pi^0e^+\nu(\gamma))}, 
     \end{equation}
     which compares radiative inclusive $K_{l3}$ for the $K_L$ and $\pi_{e3}$ decay rates.  The $K_L$
     is chosen because it is, currently, the best measured. The denominator decay rate is generally viewed as a
     theoretically pristine method for measuring $\Vud$; but, unfortunately, it is not quantitatively
     competitive with more precise determinations of that matrix element. Nevertheless, we will show that $R_V$ is competitive as a normalization
     for $K_{e3}$ and its use for the determination of $\Vus$ in much the same way but currently with about a factor of 2 less precision than $R_A$. The
     two ratios are complementary in that one tests the weak charged vector current while the other one probes the axial-vector analog.
     
\section{ Pion Beta Decay}
                      
     The branching ratio \cite{Pocanic:2003pf} for pion beta decay, $\pi^+ \to \pi^0e^+\nu(\gamma)$,  is $1.038(6)\times 10^{-8}$ where we have taken the liberty to increase the published central value by $+0.2\%$, in keeping with an
     updated \cite{Tanabashi:2018oca} normalizing $\BR(\pi^+ \to
     e^+\nu(\gamma))=1.2325(23)\times 10^{-4}$. Used in conjunction with the pion
      lifetime $26.033(5)\times 10^{-9}$ s, it
     implies the experimental decay rate
     \begin{equation}
    \Gamma(\pi^+ \to \pi^0e^+\nu(\gamma)) =0.3988(23)
    \text{ s}^{-1},  \label{pionWidth}
            \end{equation}
     which can be compared with the rather precise SM theoretical prediction \cite{Kallen:1964lxa,Sirlin:1977sv,Cirigliano:2002ng,Passera:2011ae}
       \begin{align}
       \label{eq14}
\Gamma(\pi^+ \to \pi^0 e^+\nu(\gamma)) &={G_\mu^2|\Vud|^2 m_{\pi^+}^5 \left|f_+^\pi(0)\right|^2 \over 64\pi^3} 
(1+\text{RC}_\pi) I_\pi ,
\end{align}
where \cite{Sirlin:1977sv,Kallen:1964lxa}
\begin{align}
\text{RC}_\pi &= 0.0334(10), \\
    I_\pi &= {32\over 15} \left( 1 - {\Delta \over 2m_{\pi^+}}\right)^3
   \left( \Delta \over m_{\pi^+} \right)^5 f(\epsilon,\Delta)
    = 7.376(1)\times 10^{-8},
    \\
     f(\epsilon,\Delta) & = \sqrt{ 1-\epsilon }
     \left( 1-{9\epsilon\over 2} -4\epsilon^2 \right)
     +{15\over 2}\epsilon^2 \ln{ 1 + \sqrt{ 1-\epsilon }
      \over \sqrt{ \epsilon }}
      -{3\over 7} {\Delta^2  \over (m_{\pi^+}+m_{\pi^0})^2 },
      \\
      \Delta  & = m_{\pi^+}-m_{\pi^0},\quad
      \epsilon= \nicefrac{m_e^2 }{\Delta^2},
      %{m_e^2 \over \Delta^2},
\end{align}
where the +0.0334(10) represents our estimate of the electroweak and quantum electrodynamics (QED) radiative corrections and their uncertainty \cite{Cirigliano:2002ng,Passera:2011ae,McFarlane:1994aa,Cirigliano:2003yr,Knecht:2004xr}.
Its relatively small theoretical uncertainty is justified by the good
agreement between the current algebra \cite{Sirlin:1977sv,Passera:2011ae} and chiral perturbation
theory \cite{Cirigliano:2002ng} calculations of the radiative corrections. Both include a universal short-distance 0.0234 electroweak correction combined with a 0.010(1) long-distance QED contribution.
 Solving for $\Vud$ leads to
 \begin{equation}
   \label{eq:Vud}
         |\Vud|= 0.9739(29).
 \end{equation}
      That value is in good accord with expectations from CKM unitarity; but it is not competitive with super-allowed
      nuclear beta decays which are more precise by better than an order of
      magnitude \cite{Hardy:2016vhg}.
     Even further improvement by an additional factor of 2 or 3, which appears to be possible but challenging \cite{McFarlane:1994aa} would not make pion beta decay directly competitive for
     determining $\Vud$. 
     However, we note that currently, the $\pm 0.6\%$ fractional uncertainty in the pion
     beta decay rate (see eq.~\eqref{pionWidth}) is similar to individual
     $\Klt$ rates used in the determinations of $\Vus$.  So, it
     can be used to normalize $\Klt$ decay widths without a significant
     increase in the overall uncertainty and allows the potential for further improvement.  Those features provide the basis for our following discussion. 
     
\section{ The Ratio $R_V$}

We begin by considering the $K_L{(3e)}$ partial decay width, traditionally normalized in terms of the muon lifetime derived Fermi constant, $G_\mu$. It has a form
similar  to eq.~\eqref{eq14},
\begin{equation}
     \label{eqKL}
\Gamma(K_L \to \pi^\mp e^\pm \nu(\gamma)) =
  {G_\mu^2|\Vus|^2 m_{K_L}^5 \left|f_+^K(0)\right|^2  
  \over 192\pi^3} 
(1+\text{RC}_K)
 I_K ,
\end{equation}
with \cite{Antonelli:2010yf}
    \begin{align}
\text{RC}_K &= 0.0334(20),    \\
I_K &= 
2 \left({\mpi \over \mKL}\right)^4 
\left(1 + 2 \lambda_+  { \mKL^2 + \mpi^2 \over \mpi^2}\right)
 \left[
    {\beta_m (5 \beta_m^2 - 3) E_m^4 \over \mpi^4}
    +  3 \ln{E_m (  1 + \beta_m)\over \mpi} 
\right] 
    - {64 \lambda_+ \beta_m^5  E_m^5 \over 5 \mpi^2 \mKL^3}
    \nonumber \\
    &= 0.15455(15) \text{ and } 
    \beta_m  =\sqrt{1  -\left( \nicefrac{\mpi}{E_m}\right)^2},
\end{align}
where $E_m = 0.26838$ GeV is the maximum pion energy in the $K_L$ rest frame and
$\lambda_+= 0.0282(4)$ parametrizes the average linear energy dependence (slope) of the form
factor $f_+^K$ \cite{Tanabashi:2018oca}. Although we employ only the
average  $\lambda_+$ term in the form factor expansion for illustrative simplicity,
the $I_K$ value found is very close, within (roughly) 1 $\sigma$, to the result for
$I_K$ obtained keeping higher order terms in the form factor  expansion that is fit
to the physical decay spectrum. Alternatively, $I_K$ can be obtained using a
dispersive approach. We actually employ the updated dispersive value $I_K$
=0.15470(15) \cite{passemar:2019aa} in our following analysis. The radiative corrections (RC$_{\pi,K}$) in
eq.~\eqref{eq14} and eq.~\eqref{eqKL} are to a very good approximation equal in
magnitude and cancel (up to the uncertainties) in the ratio, $R_V$, defined in eq.~\eqref{eq:RV},
\begin{equation}
   { 1+\text{RC}_K \over 1+\text{RC}_\pi}\simeq 1+\text{RC}_K - \text{RC}_\pi = 1.000(2)_K(1)_\pi. \label{new23}
\end{equation}
Cancellation includes common uncertainties in the short distance
electroweak and QED radiative corrections. Short-distance cancellation is expected; but
the QED long-distance 0.010(2) cancellation with 0.010(1)
appears to be accidental. Long distance SM effects for
other $K$ decay modes will differ somewhat, numerically
\cite{Cirigliano:2008wn,Antonelli:2010yf} from $\text{RC}_K =
0.0334(20)$ and do not fully cancel. Their remainders and correlated
theory uncertainties are included as part of the $K$ contribution
when we average over all $\Klt$ decay modes. The current underlying
radiative correction uncertainty is estimated to be about 0.2\% at the decay
rate level 
\cite{Cirigliano:2008wn}.
It represents a theoretical limitation for future $\Klt$ extractions of
$\Vus$. We note, however, that further reduction in the radiative corrections uncertainty may be possible
using the current algebra formalism \cite{Sirlin:1977sv} for $K_{l3}$ decays as a check on chiral perturbation theory, an approach recently
advocated by Seng, Galviz and Mei{\ss}ner \cite{Seng:2019lxf}. It could, in principle, allow one to refine the calculation of
$\text{RC}_{\pi}-\text{RC}_K$
and compare with the values we have taken from \cite{Sirlin:1977sv,Passera:2011ae,Cirigliano:2008wn,Antonelli:2010yf}.  Given the current excellent agreement among the different $\Klt$ decay modes, we don't expect major changes; but better precision along with a check of the long distance QED corrections would be welcome.
At present, experimental improvement in $R_V$ up to about a factor
of 3 can be envisioned
before confronting theoretical uncertainty in the radiative corrections. 

The usual method for extracting $\Vus$ is to compare
the $K_L{(3e)}$ partial width theory with experiment to obtain the constraint $f_+^K(0)\Vus=0.2165(6)$. Employing a lattice gauge theory
calculation for the form factor,  is then used to determine
$\Vus$ at about the 0.3\% level.  Its value depends on the form factor as well as factorized electroweak short and QED long-distance radiative corrections along with the Fermi constant, all of theoretical origin and assumed to be consistent with the definitions of $\Vus$ and $f_+^K(0)$ employed. Next, as an alternative, we  normalize relative to pion beta decay which provides a different perspective
on testing CKM unitarity and a means to search for the presence of ``new physics''. Of course, a similar exercise can be carried out for any of the $\Klt$  neutral and charged kaon decay modes.
Then, the results (including correlated errors) can be averaged to somewhat reduce the uncertainties.

The ratio $R_V$, defined in eq.~\eqref{eq:RV}, has the current experimental value
\begin{equation}
    R_V^{\text{exp}}= {\tau_\pi \times \text{BR}(K_L \to\pi^\pm e^\mp\nu(\gamma)) \over \tau_{K_L} \times \text{BR}(\pi^+ \to \pi^0 e^+\nu(\gamma))}= { 26.033(5)\,\text{ns}\times 0.4056(9) \over 51.16(21)\,\text{ns} \times 1.038(6)\times 10^{-8}}=1.9884(115)(93)\times 10^7, \label{eq:RVexp}
\end{equation}
    where the first uncertainty stems from the pion partial width and
    the second from the $K_L$ lifetime and branching ratio. The latter
    can be reduced by roughly a factor $\nicefrac{2}{3}$  by averaging over all $\Klt$
    partial widths after accounting for differences in phase space, QED corrections, particle masses, a second form factor for muon modes and in the case of charged kaons strong isospin breaking \cite{Cirigliano:2001mk,Cirigliano:2007zz,Cirigliano:2008wn}. 
    
    The experimental value of eq.~\eqref{eq:RVexp} should be compared with the SM prediction \cite{Cirigliano:2001mk,Cirigliano:2007zz,Antonelli:2010yf,Moulson:2017ive,moulson:2019aa,passemar:2019aa}
    \begin{align}
    R_V^{\text{theory}}&={1\over 3}\left({m_{K^0}\over
                         m_{\pi^+}}\right)^5\left(   { f_+^K(0)\over
                         f_+^\pi(0) }{\Vus\over
                         \Vud}\right)^2\times{I_K\over I_\pi}\times 1.000(2).
\end{align}
     Equating theory and experiment leads to
     \begin{equation}
       \label{eq:20}
      { f_+^K(0)\over f_+^\pi(0) } {\Vus\over \Vud} = 0.22220(64)(58) ,
                \end{equation}
      with the first and second error coming from the pion and kaon decay modes respectively. The second error (58) in eq.~\eqref{eq:20} 
     includes  uncertainties from $I_K$ and RC$_{K}$. Together they
     contribute about (22) to the (58).
     The radiative corrections (modulo uncertainties) (RC$_{\pi,K}$) and Fermi constant have cancelled out in the ratio,  $I_K$ and $I_\pi$ are phase space integrals and $f_+^{K}(0)$ is the $K\pi$ vector transition form
     factor, which is 1 up to second order in SU(3) flavor breaking \cite{Ademollo:1964sr}
     for the neutral kaon. (The charged kaon form factor requires \cite{Leutwyler:1984je}
     about a 2.8(3)\% increase to fully account 
     for its difference with the neutral kaon case. 
     That shift is consistent with expectations due to  
     strong isospin breaking resulting from pion-eta mixing.)
     The pion form factor, $f_+^\pi(0)$, is essentially 1 in the
     SU(2) flavor limit \cite{Behrends:1960nf}.  The deviation from 1 is $\order{10^{-5}}$ and can be usually ignored. However, we retain the form factor ratio $f_+^K(0)/f_+^\pi(0)$ in our
     discussion, since for some lattice calculations that ratio may provide a means
     for extraneous lattice artifacts, such as finite size effects, to cancel.

     \begin{table}[htb]
\centering
\begin{tabular}{l@{\hspace*{10mm}}l@{\hspace*{10mm}}l}
\hline 
\hline
\vspace*{1mm}
Decay &
$  { f_+^K(0) |\Vus|\over f_+^\pi(0) |\Vud| }$ & 
$  { f_+^K(0) \over f_+^\pi(0) }$  \\
\hline
 $K_L(e3)$ & 0.22220(64)(58) & 0.9606(28)(19)(25) \\
 $K_L(\mu 3)$ & 0.22250(64)(64) & 0.9619(28)(19)(28) \\
 $K_S(e3)$ & 0.22138(64)(134) & 0.9571(28)(19)(52) \\
 $K^\pm (e3)$ & 0.22220(64)(86) & 0.9606(28)(19)(37) \\
 $K^\pm (\mu 3)$ & 0.22200(64)(111) & 0.9602(28)(19)(48) \\
\hline 
Average & 0.22223(64)(40) & 0.9607(28)(19)(18) \\
\hline 
\hline 
\end{tabular}
\caption{ $\Klt$ results from five decay modes with approximate errors, weighted average (including some correlated theory uncertainties) for ${ f_+^K(0) |\Vus| / f_+^\pi(0) |\Vud| }$, \cite{Cirigliano:2008wn,Antonelli:2010yf} based in part on the updated results in \cite{moulson:2019aa,passemar:2019aa}. Also shown are the individual $  { f_+^K(0) / f_+^\pi(0) }$ values and their average for ${|\Vus| / |\Vud| }=0.23131(45)$ (see eq.~\eqref{eq:X}).}
\label{table1}
\end{table}   
     
     Equating $R_V$ experiment and theory followed by weighted averaging over
     all $\Klt$ modes, allowing for correlated uncertainties, as shown in Table \ref{table1}, following the literature cited,
     our analysis roughly corresponds to rescaling the carefully
     studied 5 known ${ f_+^K(0) |\Vus|}$ values. It leads to the
     average      
     \begin{equation}
       \label{eq:21}
      {f_+^K(0)\Vus\over f_+^\pi(0)\Vud}=0.22223(64)(40), 
            \end{equation}
      where the central value has remained nearly unchanged; 
      but, the kaon dependent uncertainty has been reduced by 
      about a factor of $\nicefrac{2}{3}$. We note that the  $\chi^2$/degree of freedom for the
      five $\Klt$ modes was found \cite{moulson:2019aa,passemar:2019aa} to be an acceptable 0.98/4; so, it is not necessary 
      to scale up the error. The goodness of the fit also helps validate the
      relative magnitude of the radiative corrections applied to the different $K$ decay modes.

      Requiring $R_V$ and $R_A$ to have
      the same $\Vus/\Vud=0.23131(45)$  
      within errors implies $f_+^K(0)/f_+^\pi(0)=0.9607(38)$.  For $f_+^\pi(0)=1$ the implied $f_+^K(0)=0.9607(38)$ is in tension with the prevailing \cite{Bazavov:2018kjg} lattice \QCD result of roughly $0.970(2)$ for 2+1+1
      quark flavors by about 2.2$\sigma$.  That discrepancy can also be illustrated
      by inserting $f_+^K(0)/f_+^\pi(0)=0.970(2)$ in eq.~\eqref{eq:21} which leads to
      $\Vus/\Vud=0.22910(91)$. It differs from eq.~\eqref{eq:X} by a related 2.2$\sigma$.
      If we only used $K_L(e3)$ in our comparison,
      the 2.2$\sigma$ difference would have been
      2.0$\sigma$. Averaging over the five decay modes
      has not had a dramatic effect. It does, however,
      demonstrate the consistency of our experimental
      and theoretical input.
      
      The roughly 2$\sigma$ discrepancy was already observed 
      some time ago in global $K$ decay fits \cite{Antonelli:2010yf}. Currently, averaging over the five  kaon beta decay modes indicates
      about a 2.6$\sigma$ discrepancy with the unitarity based  $R_A$ = 0.23131(45) constraint for $\Vus$. An alternative interpretation of the apparent lattice value discrepancy is the need for an additional roughly $-0.01$ electromagnetic radiative correction relating electromagnetically free $f_+^K(0)=0.97$
      obtained from pure lattice 
      QCD with the form factor appropriate for the decay rate formula in eq.~\eqref{eqKL}.
      Such an effect would increase the central value of $\Vus$ to 0.2253, consistent with $R_A$ and CKM unitarity.
      If instead new physics is responsible for the discrepancy, it is less likely due to 
      short-distance electroweak radiative corrections, muon lifetime $G_\mu$
      normalization or some other effect common to numerator and denominator which cancels in the $R_V$ ratio. We note that a similar discrepancy occurs
      in 2+1 flavor lattice gauge theories which predict $f_+^K(0)= 0.968$; supporting the interpretation that a reduction in the lattice form factor by about $-0.01$  may be the most likely route to CKM unitarity. To
      resolve this situation, additional lattice studies and scrutiny are needed.
      
      We also note that employing $\Vus/\Vud=0.22910(102)$ suggested from $R_V$, lattice $f_+^K(0)=0.970(2)$ plus three-generation unitarity implies
      $\Vud=0.97474(22)$ which exceeds eq.~\eqref{eq:11}. That makes it more difficult to
      reconcile with the current super-allowed nuclear beta decay discrepancy.
      
      We averaged over all five $\Klt$ decay modes to reduce the second
      error in eq.~\eqref{eq:20} by roughly a factor of $\nicefrac{2}{3}$. Future improvements in kaon measurements are expected \cite{Cirigliano:2007zz,Antonelli:2010yf} to
      further reduce that part of the uncertainty by another
      $\nicefrac{2}{3}$, leaving the partial width of pion beta decay
      as the dominant uncertainty in the error budget by about a factor of 2.5. Improving the experimental pion partial beta decay width by a factor of 2 to 3 would bring the overall
      experimental error budget for $R_V$ down by roughly
      a factor of 2. At that level, averaging over kaon modes becomes more important. Such a reduction in the $\Vus/\Vud$ uncertainty 
      derived from $R_V$ along with a similar improvement in $R_A$ will together strongly restrict or provide evidence for the existence of
      ``new physics" at potentially high significance. The latter scenario would be more likely if the $\Vud$ from super-allowed beta decays continues to show a deviation from CKM unitarity. Neutron lifetime and decay asymmetry precision measurements should also help resolve the $\Vud$ problem \cite{Czarnecki:2018okw,Czarnecki:2019mwq}.  Indeed, the larger radiative corrections found in \cite{Seng:2018yzq,Czarnecki:2019mwq} combined with a unitarity favored 0.97428 for
      $|\Vud|$ and 1.2762 for $g_A$ predict a neutron lifetime of about
      878 s with a small uncertainty.
      
      More precise experimental measurements are clearly needed to reconcile CKM unitarity or unveil evidence for ``new physics." Our study of pion beta decay and the utility of $R_V$ will hopefully reinvigorate interest in that experimental effort.  More specifically,
      on the basis of its complementary role, we advocate
      a new experiment on pion beta decay designed to improve measurement of that rare branching ratio by an overall factor of 2 to 3 \cite{Pocanic:2003pf}. At that level, combined with anticipated $K_{l3}$ improvements, it could provide the best determination of $\Vus$ or at least a consistency check on lattice calculations. In addition, we encourage the lattice gauge theory community to examine the possibility of a reduced uncertainty to 0.001 and check for a difference in the definition of the lattice $f_+^K(0)$ and the form factor as defined by $R_A$

     Note Added: 
After our paper was submitted for publication, a preprint
by the PACS Collaboration appeared \cite{Kakazu:2019ltq}. It finds
for a large lattice with 2+1 quark flavors, $f_+^K(0)=0.960(5)$. Also, a paper 
by X. Feng {\em et al.}~\cite{Feng:2020zdc} was posted  on the arXiv. It confirms our our calculation of the pion beta 
decay rate using a novel lattice approach and reduces the theoretical uncertainty 
by a factor of 3, strengthening our  result 
and its call for an improved experimental measurement of that rare decay mode.

Acknowledgement:\\ 
The work of A.~C.~was supported
by the Natural Sciences and Engineering Research Council of
Canada. The work of W.~J.~M.~was supported by the U.S. Department of
Energy under grant DE-SC0012704. The work of A.~S.~was supported in
part by the National Science Foundation under Grant PHY-1620039.

\def\disclaim{1}
\ifnum \disclaim=1
Notice: This manuscript has been co-authored by employees of
Brookhaven Science Associates, LLC under Contract No. DE-SC0012704
with the U.S. Department of Energy. The publisher by accepting the
manuscript for publication acknowledges that the United States
Government retains a non-exclusive, paid-up, irrevocable, world-wide
license to publish or reproduce the published form of this manuscript,
or allow others to do so, for United States Government purposes.  This
preprint is intended for publication in a journal or
proceedings.
Since changes may be made before publication,
 it may not be cited or reproduced without the author's permission.
\subsection*{DISCLAIMER}

This report was prepared as an account of work sponsored by an agency
of the United States Government. Neither the United States Government
nor any agency thereof, nor any of their employees, nor any of their
contractors, subcontractors, or their employees, makes any warranty,
express or implied, or assumes any legal liability or responsibility
for the accuracy, completeness, or any third party's use or the
results of such use of any information, apparatus, product, or process
disclosed, or represents that its use would not infringe privately
owned rights. Reference herein to any specific commercial product,
process, or service by trade name, trademark, manufacturer, or
otherwise, does not necessarily constitute or imply its endorsement,
recommendation, or favoring by the United States Government or any
agency thereof or its contractors or subcontractors. The views and
opinions of authors expressed herein do not necessarily state or
reflect those of the United States Government or any agency thereof.
\fi

%\bibliographystyle{/Users/czar/Dropbox/pro/Tables/Archive/Bibtex/andrzej}
%\bibliography{/Users/czar/Dropbox/pro/Tables/Archive/ac}
%\bibliographystyle{andrzej}
%\bibliography{ac}

\end{document}